# A Survey on Cloud Computing Security

Hero Modares (Corresponding author)

Department Of Computer Systems & Technology,

University Of Malaya, 50603 Kuala Lumpur, Malaysia

Tel: (+60)-176-666-4612    E-mail: hero.modares@gmail.com

Rosli Salleh

Department Of Computer Systems & Technology,

University Of Malaya, 50603 Kuala Lumpur, Malaysia

E-mail: rosli_salleh@um.edu.my

Amirhosein Moravejosharieh

Department Of Computer Systems & Technology,

University Of Malaya, 50603 Kuala Lumpur, Malaysia

E-mail: amirhosein.moravej@yahoo.com

Hassan Keshavarz

Department Of Computer Systems & Technology,

University Of Malaya, 50603 Kuala Lumpur, Malaysia

E-mail: keshavarz_hassan@IEEE.org

Majid Talebi Shahgoli

Asia Pacific University College of Technology and Innovation,

Kuala Lumpur, Malaysia

Email: Majid.talebi@gmail.com

**Abstract**

Computation encounter the new approach of cloud computing which maybe keeps the world and possibly can prepare all the human's necessities. In other words, cloud computing is the subsequent regular step in the evolution of on-demand information technology services and products. The Cloud is a metaphor for the Internet and is a concept for the covered complicated infrastructure; it also depends on sketching in computer network diagrams. In this paper we will focus on concept of cloud computing, cloud deployment models, cloud security challenges encryption and data protection, privacy and security and data management and movement from grid to cloud.

**1. Introduction**

Development of many acquired technologies and advances to computing into something different are summarized in a "Cloud Computing" or "cloud" term which detaches basic infrastructure from application and information resources, and the mechanisms used to deliver them. Cloud improves collaboration, activity, scaling, and availability, and causes to cost reduction under the aegis of optimized and efficient computing. Distinctively, cloud depicts the usage of a collection of services, applications, information, and infrastructure consisted of pools of compute, network, information, and storage resources. Arranging, preparation, accomplishing and decommissioning and also scaling up or down are immediately performed on mentioned constituents and they are supplied for an on-demand utility-like model of allocation and consumption. Similarity and diversity of cloud from present models of computing are so puzzling, considering architectural viewpoint; and also there are much questions about their effects on the organizational, operational, and technological functions to network and information security practices. Today academicians, architects, engineers, developers, managers, and consumers all have their own specific description for cloud [1, 2, 5]. In the next session we will focus on concept of cloud computing.

*1.1What is Cloud Computing?*

Cloud computing cause to acquire shared resources and common infrastructure at ease, and affording on-demand services over the network to accomplish operations that come upon business needs variation. Typically, the last user does not have knowledge about location of accessible physical resources and tools. Evolving, utilizing and managing their applications 'on the cloud', which involves virtualization of resources that keeps and directs itself are conducted by prepared facilities to users.

Work of the scientists at the U.S. National Institute of Standards and Technology (NIST) and their descriptions for cloud computing disclosed after the promoted definition of the Cloud Security Alliance's guidance in earlier version. We have chosen to cooperate with the NIST Working Definition of cloud computing, as it is well accepted on the whole. So coherence and unanimity around a common language is a consequence of it and we can pinpoint on applicable cases rather than semantic nuance. Accordingly all organizations in the world are broadly used and applied this guide, but NIST is a U.S. government organization and the choosing of this source model should not be taken to suggest the exclusion of other points of view or geographies. Five essential qualities, three cloud service models, and four cloud deployment models explain cloud computing in NIST. Table. 1 is a schematic form of them which are explained below [5, 8].

A.  Software as a Service (SaaS), sometimes introduced as Service or Application Clouds. In SaaS, particular cloud abilities are suggested some performances for particular business functions and business activities, i.e. they provide applications / services by a cloud infrastructure or platform instead of providing cloud features them. Also a cloud suggested the kind of standard application software practicality. For examples: Google Docs, Salesforce CRM, SAP Business by Design.

B.  In general, Cloud Computing is not limited to Infrastructure / Platform / Software as Service systems, even though it provides improved abilities to enable these systems. I/P/SaaS can be regarded as specific "usage patterns" for cloud systems which refer to models already approached by Grid, Web Services etc. Cloud systems are an encouraging way to accomplish these models and develop them further [5, 11, 12].

C.  Platform as a Service (PaaS), a platform can prepare computational resources based on developing and hosting characteristics of applications and services. Usually dedicated APIs are used in PaaS to manage the behavior of a server hosting engine which performs and replicates the performance based on client demands. Each provider reveals his / her own API according to the respective main capabilities, so applications just can improved for one specific cloud provider and cannot be transferred to another cloud host – Although there are  efforts to broaden common programming models with cloud abilities [5, 11, 12]. For examples: Force.com, Google App Engine, Windows Azure (Platform).

D.  Infrastructure as a Service (IaaS) also introduced to as Resource Clouds, (controlled and scalable) resources are the provided services for the user – in other words, they basically provide improved virtualization abilities. Therefore different resources may be provided through a service interface:

E.  Data & Storage Clouds manage certain access to data of potentially dynamic size, comparing resource

operation with access requisites and / or quality definition. For examples: Amazon S3, SQL Azure.

F. Compute Clouds prepare access to computational resources, i.e. CPUs. Such low-level resources cannot really be utilized on their own yet and are generally disclosed as part of a "virtualized environment" (not to be combined with PaaS). So Compute Cloud Providers basically afford the ability to provide computing resources (i.e. raw access to resources contrary to PaaS which afford full software stacks to improve and create applications), typically virtualized, wherein to perform cloudified services and applications. IaaS (Infrastructure as a Service) affords extra abilities over a simple compute service for examples: Amazon EC2, Zimory, Elastichosts [5, 11, 12].

## 2. Cloud Deployment Models

Disregarding of the employed service model (SaaS, PaaS, or IaaS), four deployment models are introduced for cloud services, with derivative changes that describe specific requirements [5, 12, 14]:

A. Public Cloud: Infrastructure of the public cloud is arranged to be accessible for the general public or a large industry group and also it is owned by an organization selling cloud services. This environment can be used publically so it includes individuals, partnerships and other kinds of organizations. Public clouds are generally managed by third parties or vendors over the Internet, and services expenditures are on pay-per-use basis. Provider clouds also refer to this type. Business models like SaaS (Software-as-a-Service) and public clouds supplement each other and qualify companies to enhance shared IT resources and services.

Advantages

• Development, deployment and management of business applications at affordable expenses are extensively applied in Public clouds.

• Organizations can quickly convey highly scalable and reliable applications at more affordable expenses.

Limitations

• Regarding the security is so important in public clouds.

B. Private Cloud: A single organization just makes use of private cloud infrastructure. It is probably administered by the organization or a third party, and may be seen as on-premise or off-premise cases. This cloud computing environment lies in an organization limits and is used widely for the organization's advantage and also introduced as internal clouds. They are firstly made by IT departments within enterprises. It is good to mention that enterprises look for optimizing employment of infrastructure resources within the enterprise by supplying the infrastructure with applications using the concepts of grid and virtualization [5, 11, 14].

Advantages

• Average server utilization are improved and also low-cost servers and hardware are permitted to use for providing higher capabilities; consequently costs are reduced that in other respects, a greater number of servers would require more.

• High levels of automation cause decrease in operations costs and managerial overheads

Limitations

• IT teams may have to spend in buying, creating and controlling the clouds independently in the organization.

C. Community Clouds: Generally local infrastructure is the operation environment for cloud systems, i.e. public clouds providers suggest their own infrastructure to clients. In spite of the fact that the provider could resell the other provider's infrastructures, clouds do not collect infrastructures to establish larger, cross-boundary structures. Community clouds considered as benefit for specific smaller SMEs in which several entities accord with their related (smaller) infrastructure. Public clouds or dedicated resource infrastructures can also be collected by Community clouds. By this means, private and public community clouds are determinable. For instance smaller organizations may aggregate just to share their resources for creating a private community cloud. Otherwise resellers such as Zimory may share cloud resources from several providers and resell them. Community Clouds are still just an image, although some signs as Zimory [6] and RightScale [9] are already exist for such development. Community clouds and GRIDs technology show some characters in common [10, 11].

D. Hybrid Cloud: Combination of two or more clouds (private, community, or public) make the cloud infrastructure which remain as unique entities however normalized or characteristic technology that authorizes information and application portability attached them to each other (e.g., cloud bursting for load-balancing between clouds). This is a composition of both private (internal) and public (external) cloud computing environments [5, 12].

**3. Cloud Deployment Models**

In a short view, the security necessities for providers of the cloud computing would seem to be like the traditional datacenters — employ a strong network security circumference and remove the bad guys. However, physical separation and hardware-based security cannot keep safe against attacks between virtual machines on the same server. Virtual machines from several organizations will require to be co-placed on the same physical resources to benefit the cloud computing providers from efficiencies of virtualization. Some of the basic concerns plan that businesses should inform of them when designing their cloud computing deployments are pointed below [3, 4, 11, 13, 15].

**A. Administrative Access To Servers And Applications**

Offering "self-service" access to computing power is one of the most significant properties of cloud computing, most likely via the Internet. Administrative access to servers is managed and limited to direct or on-premise connections in traditional datacenters. Now this administrative access in cloud computing must be lead through the Internet, spreading disclosure and risk. To limit administrative access and also check this access to keep visibility of system control variation is intensely important [15].

**B. Dynamic Virtual Machines: VM State And Sprawl**

Virtual machines are dynamic. They can be reverted to previous stages, stopped and restarted at speed, rather easy. They can also be easily cloned and moved between physical servers continuously. Obtaining and preserving uniform security considering dynamic machines' characters and their potential for VM sprawl is complicated. Weaknesses or configuration errors may be distributed unintentionally. Keeping the auditable record of the security state of a virtual machine at any requested time is hard too. It is necessary to be able to confirm the security state of a system in cloud computing environments, without regards to its location or closeness to other, capability vulnerable virtual machines [15].

**C. Vulnerability Exploits And VM-To-VM Attacks**

The same operating systems, enterprise and web applications are used for both cloud computing servers and localized virtual machines and physical servers. Exploitation of vulnerabilities distantly by attacker or malware potential in these systems and applications is a serious warning to virtualized cloud computing environments. Besides, when several virtual machines co-located cause to enhance in the attack surface and risk of VM-to-VM compromise. Without consideration of the location of the VM within the virtualized cloud environment, it is important to know that intrusion recognizing and restraint systems need to be able to recognize spiteful activity at the virtual-machine level [15].

### D. Data Integrity: Co-Location, Compromise And Theft

Report of 2008 Data Breach Investigations which are conducted by Verizon Business Risk Team, showed that 59% of data breaches are the consequence of hacking and intrusions. Dedicated resources in comparisons with shared resources are seemed to be more protected, so the expectation of greater attack surface in fully or partially shared cloud environments result in increased risk. Enterprises require assurance and auditable verification that cloud resources are not being changed with nor compromised, specifically when existing in shared physical infrastructure. Operating system and application files and activities should be recorded [15].

### E. Encryption And Data Protection

Requirements for the use of encryption to preserve crucial data—such as cardholder data and personally distinguishable information (PII)— and also to acquire compliance or secure harbor in breaching to be found in many procedures and standards such as the PCI DSS and HIPAA. These necessities are heightened by the multi-tenant character of cloud, and causes unique challenges with the availability and security of encryption credentials used to guarantee data safety [15].

### 4. Data Management

Expanding of the total accessible data on the web, as well as the throughput generated by applications, sensors are occurred faster than storage and in particular bandwidth function. Developed means of administrating and structuring the extent of data will be essential to handle future requirements since a strong inclination exist to host more and more public data sets in cloud infrastructure, consequently such meaning requirements should be supplied by storage clouds in order to keep accessibility of information and accordingly introduce quality requisites etc. In addition to data size, uniformity maintenance also poses a more importantly problem for cloud systems, especially when scaling up. Probably information shared to some extent or totally between tenants, i.e. supplying uniformity over a potentially vast number of data instances becomes more and more important and complex as the whole database is replicated or actually a subset is prone to simultaneous access such as state information. Efforts in providing transactional assurance for software stacks absolutely is one of major analysis gap in this area (e.g. multi-tier architectures as SAP NetWeaver, Microsoft .NET or IBM WebSphere) and also it provisions large scalability (100s of nodes) without resorting to data separation or moderate uniformity (such as ultimate uniformity) [11].

Now, partitioning and scattering of data occurs rather uncontrollably so it leads potentially to coincide with legislation moreover proficiency issues and (re)integration problems. More control potentials are needed over scattering in the infrastructure for compensating this problem, and these controls allow for environment analysis and QoS accomplishment a feature that is scarcely inscribed by commercial and / or research functions until.

Various data sources cause almost all data in the web to be unstructured and heterogeneous so new forms of explanation for perceptive separation and information operation are needed. Furthermore, uniformity maintenance procedures probably change between data formats, in this case it is just compostable by maintaining meta-information about operation and structure. By considering the single cloud systems and suitableness structures of them, it appears that transferring data (and / or services) between these infrastructures is occasionally difficult and indeed need to develop new standards and secure long period interoperability. Study on the "eXternal Data Representation" (XDR) standard will have an important role for systems which coupled insecurely. Multiple tenants potentially have cloud resources in common– in other words it is divided between them not only to storage (and CPUs, see below), but also to data (where e.g. a database is shared between multiple users) so variations are found at different locations and also in a simultaneous style. This requirements developed means to handle multi-tenancy in scattered data systems. Systems with traditional data management stop working because of great numbers of nodes – even if grouped in a cloud. Delay in accessing disks shows when keeping of basic part of the system universal mode is essential classical transaction controlling (two-phase performance) not to be sustainable. Effectiveness efforts combine the problem requiring cache coherency across a vast number of nodes. Present clouds generally use either concentrated Storage Area Networks, unshared local disk or cluster file-systems, therefore products storage are not able to combine with cloud storage presently at ease, although Live Mesh let for local storage synchronization in / with the cloud by this time. Genuine operation manner with considering the file and data access in cloud systems should be evaluated more cautiously to introduce these issues. Just few of these studies exist now [7], but recognizing the typical scattering, access, uniformity etc. which are the necessities of the individual operation cases are performed accompanied by matching information.

**5. Privacy & Security**

The multi-tenant resources –shared resources- and the potentially undefined location of resources eventuate to concern of data protection and other potential security holes that some issues regarding codification and data scattering are used for solving them. Specifically outsourcing issues critically require perceptive data or secured applications. In some operation cases, industrial spying is able to perform by the information that a certain industry is using the infrastructure at all in other words that kind of information is sufficient for espionage[3, 11]. Extra issues employ through the specifics of cloud systems while necessary security features are introduced by most devices, additional ones particularly associated to the replication and scattering of data in potentially global resource infrastructures. Information should be secured in a form that introduces legislative issues considering data location; meanwhile it should be controllable by the system too.

Moreover, the frequent usages of cloud systems and the diversity of cloud types indicate several security models and necessities by the user. Classical authentication models when the computational image hosts services that are accessible to users, may not be enough to recognize the Aggregators / Vendors from the real User, specifically in IaaS cloud systems. Where the cloud systems are collected or resoled, the compound of security procedures probably don't just result in problems of compatibility, they also cause to the user scattering the model because of insight absence. Totally, new security managing models & operations are necessary for providing particular issues which are consequence of the cloud model [3, 4].

**6. Movement from Grid to Cloud**

Although in general Europe progress slower than the US by considering the industrial cloud movement and also it seems that Europe has less resource infrastructure available, but a relatively large group of Grid dealers and uptakers are exist in Europe. The mighty similarities between the business motivation of Grid dealers and Cloud providers, and also common requirements towards the infrastructure cause relatively easy for current (European) Grid vendors to proceed towards cloud providing (including protecting devices and middleware) and already being assumed by companies such as GridSystems. European market competitor can be regarded "ready" for acquiring a cloud service offering, especially in the Grid area. For administrating that step, it must become apparent to them [11]:

a) How their business can be developed?

b) Why pursuing this movement are necessary for all users and finally?

c) How this can be accomplished and how potential difficulty can be vanquished?

## 7. Conclusion

The long held reality of computing as a function is finally rising. Cloud computing is a promising computing example that is more and more popular. The cloud over the Internet provides the infrastructure required to supply services directly to customers. IBM, Google, and Microsoft as a leader in the industry, have provided their schemes in supporting cloud computing. On the other hand, the public literature that argues the research concerns in cloud computing are still insufficient. Still there are many issues have missing without an answer and to be sure the majority important is security. Social feature is one of the other phases of the cloud which is missed. The outlook of the Cloud computing involves work out with virtual representations without the physical existence of software or hardware. For example, programmers can design their applications using a web-based platform. Users can manage their businesses with simple to modify online software. There are various profits from switching over to Cloud computing. The first advantage is that capital expenditure costs are decreased as new systems and protection costs are decreased by way of a pay-as-you-go methodology. Developers are declined of their concerns concerning interoperability. As Cloud computing grows to be more common, difficulty regarding official problems start to come up therefore useful polices are required on both the local level and international.

On the other hand, the upcoming of Cloud computing is not all that light, if fact it could be extremely Cloudy except official risks and possible security are traded with and a prosperous business model is created. There are risks and hard security tasks related to Cloud computing which essentially turn around the issues of compliance, accessibility of bandwidth, global execution, IP violation, exploit of client data, location of client data, transparency and most of all the question in case of legal battle. Businesses today require certainty and assurance to handle any service provider; they also require the pliability to exploit services for their business requirements of changeable levels of legal challenges.

Table 1. NIST Visual Model of Cloud Computing Definition

| Broad Network Access | Rapid Elasticity | Measured Service | On-Demand Self-Service | Essential Characteristics |
|---|---|---|---|---|
| Resource Pooling | | | | |
| Software as a Service (SaaS) | Platform as a service (PaaS) | Infrastructure as a Service (IaaS) | | Service Models |
| Public | Private | Hybrid | Community | Deployment Models |